\documentclass[conference]{IEEEtran}

\usepackage{multicol}

\usepackage{color}
\usepackage{xcolor}
\usepackage{soul}
\usepackage{amsmath}
\usepackage{amssymb}
\usepackage{graphics}
\usepackage{setspace}
\usepackage{cite}
\usepackage{latexsym}
\usepackage{float}
\usepackage{epsfig}
\usepackage{multirow}
\usepackage{cite,cases,url}
\usepackage{graphicx}
\usepackage{epstopdf}
\usepackage{balance}
\usepackage{subcaption}
\usepackage{enumerate}
\usepackage{array}
\usepackage{enumitem}
\usepackage{dblfloatfix}
\usepackage[normalem]{ulem}
\useunder{\uline}{\ul}{}
\usepackage{algorithm}
\usepackage{enumitem}
\usepackage{caption}
\usepackage{subcaption}
\usepackage{tikz}
\usepackage{adjustbox}
\usepackage{etoolbox}
\usepackage[noend]{algpseudocode}
\makeatletter
\usepackage{tikzpagenodes}      
\usepackage[normalem]{ulem}
\usepackage[linesnumbered,ruled,vlined,algo2e]{algorithm2e}
\useunder{\uline}{\ul}{}

\newcolumntype{L}{>{\centering\arraybackslash}m{5cm}}
\newcolumntype{K}{>{\centering\arraybackslash}m{6cm}}
\newcolumntype{P}{>{\centering\arraybackslash}m{2.3cm}}
\newcolumntype{M}{>{\raggedright\arraybackslash}m{2cm}}
\newcolumntype{N}{>{\raggedright\arraybackslash}m{2.5cm}}

\usepackage{changepage}
\protected\edef\mathbb{%
  \unexpanded\expandafter\expandafter\expandafter{%
    \csname mathbb \endcsname
  }%
}
\usepackage{physics}

\begin{document}


\title{{DDPG Learning for Aerial RIS-Assisted MU-MISO Communications}}

\author{
\IEEEauthorblockN{Aly Sabri Abdalla and
Vuk Marojevic
}\\ \vspace{-0.5 cm}
\normalsize\IEEEauthorblockA{Department of Electrical and Computer Engineering, Mississippi State University, MS, USA\\}

Email: asa298@msstate.edu, vuk.marojevic@msstate.edu 
}
\vspace{-2 cm}
\maketitle

\begin{abstract}
This paper defines the problem of optimizing the 
downlink multi-user multiple input, single output (MU-MISO) sum-rate for  ground users served by an aerial reconfigurable intelligent surface 
(ARIS) that acts as a relay to the terrestrial base station. The 
deep deterministic policy gradient (DDPG) is proposed to calculate the optimal active beamforming matrix at the base station and the phase shifts 
of the reflecting elements at the ARIS to maximize the data rate. 
Simulation results show the superiority of the proposed scheme 
when compared to deep Q-learning (DQL) and baseline approaches.


\end{abstract}

\IEEEpeerreviewmaketitle

\section{Introduction}
\label{sec:intro}


Several studies have been conducted for enhancing the performance of wireless communications 
networks with unmanned aerial vehicles (UAVs)
~\cite{Sec, coverage, Spect}. The Third Generation Partnership Project (3GPP) has identified the challenges and solutions of emerging cellular networks incorporating UAVs~\cite{3GPP_Stnds}. It specifies the procedures for aerial user equipment and defines 
aerial relay and aerial base station to improve coverage, capacity, or provide temporary/on-demand network access. Cellular connected UAVs have been evaluated in the context of 4G and are expected to be integrated into emerging 5G networks \cite{aerpaw-VTM}.

One of the technologies driving 6G research is the reconfigurable intelligent surface (RIS). 
It allows controlling the radio frequency (RF) propagation 
for different purposes: coverage extension, multichannel communications, and physical-layer security, among others. The RIS has been suggested for indoor and outdoor settings to improve the reliability of low-power mobile communications. 
It consists of reflecting elements that can be configured by applying different phase shifts to steer the incoming radio waves into a desired direction. This can be considered as passive beamforming since the RIS does not generate nor amplify signals. 
 
Recent literature has started to investigate the potentials of RIS carried by aerial vehicles~\cite{UAVRISAly} 
using different metrics, such as energy efficiency~\cite{ARISEE}, security~\cite{ARISSEC}, and  
fairness~\cite{ARISFAIR}. These early studies 
use conventional optimization techniques for determining the phase shifts that maximize the performance for the metric of interest. 
The RIS technology was originally introduced as reflecting surfaces for building walls. More recently, such ground RISs were considered
for providing multi-user multiple input, single output (MU-MISO) wireless access. 
Such systems can be optimized by applying data driven~\cite{GRISMISORL} or conventional optimization solutions ~\cite{GRISMISOOpt}. {The enhanced communications performance of RIS-assisted MU-MISO, as shown in}~\cite{zhang}{, motivates this paper.}
{While the individual advantages of the RIS and the UAV 
for improving communications and networking have been demonstrated, combining the two technologies 
and optimizing the MU-MISO network performance under multi-user interference has yet to be 
explored. 
}
 
This paper defines the problem of an ARIS assisted cellular downlink and proposes the  
deep deterministic policy gradient (DDPG) as a model-free {off-policy} reinforcement learning (RL) technique for maximizing the sum-rate of the MU-MISO system. This is accomplished by jointly optimizing the active beamforming matrix at the ground base station (GBS) and the phase shifters of the passive reflecting elements at the ARIS. {The DDPG is proposed in this paper because of its ability to work over 
continuous action spaces
~\cite{DDPG}.} 
To the best of our knowledge, none of the existing 
studies have explored 
RL 
for optimizing the transmitted signal and reflections off an ARIS 
for 
MU-MISO communications. 
 
The rest of the paper is organized as follows. Section II provides the system model and problem formulation. Section III introduces the DDPG algorithm as our learning-based solution for 
the joint design of the GBS beamforming matrix and the ARIS phase shifts. Section IV presents the numerical analysis, comparing the proposed approach to other deep learning and baseline solutions. 
Section V provides the concluding remarks.

\section{System Model and Problem Formulation} 
\vspace{-1 mm}
\label{sec:system}

\subsection{System Model}
In this paper, we consider a MU-MISO communications system and consider the downlink, where the ARIS is deployed for relaying the signals from the GBS to a number of active ground terminals (GTs). This is depicted in Fig. 1.  
The 
GBS is 
equipped with $M$ antennas and located at $\vb*{q_T}=[x_t, y_t, z_t]^T$. 
The ARIS has $N$ 
reflecting elements that are controlled by a microcontroller. 
There are $K$ single antenna users, {where $K \leq M$}. We assume that the direct links between the GBS and GTs are under sever blockage due to obstruction and are therefore neglected. We also consider that all of the reflecting elements of the ARIS are allocated to one user at time to enhance the throughput of the relayed link.

For the considered 
MU-MISO communications system, the GBS leverages its $M$ antennas to transmit $K$ independent data streams 
to the ARIS at 
$\vb*{q_R}=[x_r, y_r, z_r]^T$,  
where each data stream is intended for one of the $K$ 
GTs. 
The simultaneously received signals at the ARIS are 
reflected 
via the reflecting elements toward the $K$ static GTs whose locations are $\vb*{q_k}=[x_k, y_k]^T$ for $k \in [1,2, \cdots, K]$. 

 \begin{figure}[t]
     \centering
     \includegraphics[width=0.4\textwidth]{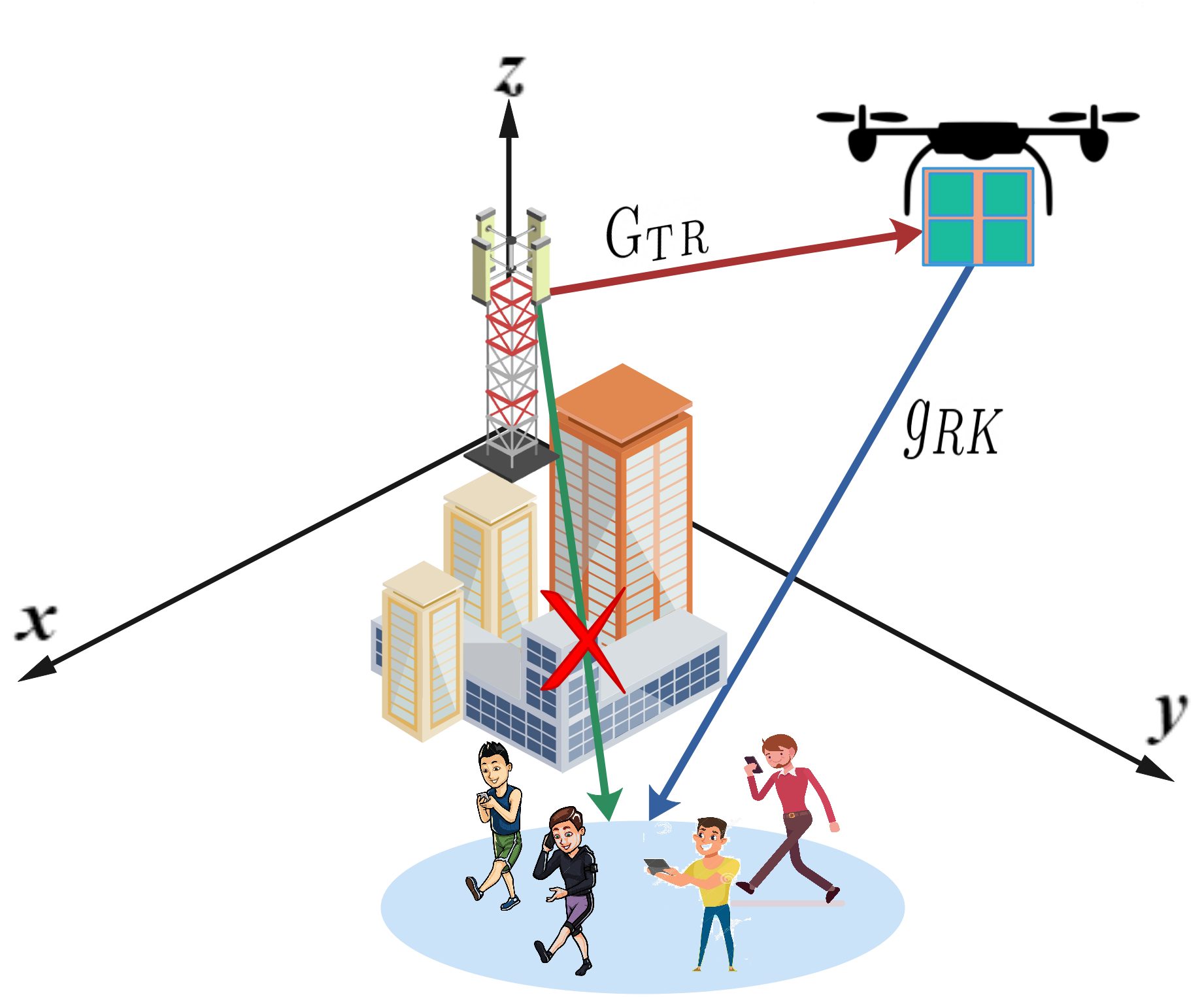}
     \caption{The ARIS-assisted MU-MISO scenario considered in this paper.}
     \vspace{-5mm}
         \label{fig:system}
 \end{figure}
 
The multiple input-multiple output (MIMO) channel between the $M$ antennas of the GBS $T$ and the $N$ reflecting elements at the ARIS $R$ is denoted as $\vb*{G_{TR}} \in \mathbb{C}^{N \times M}$ and the MISO channel between the ARIS $U$ and the $k^{th}$ GT is defined as $\vb*{g^k_{RK}} \in \mathbb{C}^{N \times 1}$. 
We assume that 
the 
channel state information (CSI) 
is perfectly known at the GBS and reported back to the ARIS over a dedicated control channel. 

The ARIS forms a uniform linear array (ULA) of $N$ reflecting elements, as in~\cite{zhang}.  
The phase shift array 
is denoted as $\vb*{\phi} \in \mathbb{C}^{N\times N}, \vb*{\phi}=diag\{e^{j\theta_{1}}, e^{j\theta_{2}}, \cdots, e^{j\theta_{N}}\}$, where $\theta_{n} \in [0, 2\pi], n \in [1,2,...,N]$ 
is the phase of the $n^{th}$ reflecting element.

The signal-to-interference plus noise ratio (SINR) at the $k^{th}$ GT can be written as 
\begin{equation}
        \gamma_{k} =   \frac{\mid \vb*{g}_{Rk} \vb*{\phi} \ \vb*{G}^{TR} \ \vb*{w}_{k} \mid^2}{\sum\limits_{i\neq k}^K \mid \vb*{g}_{Rk} \vb*{\phi} \ \vb*{G}^{TR} \vb*{w}_{i} \mid^2 +  \sigma^2_{k}} 
    \label{eq:xs}, 
\end{equation}
where the first term in the denominator is the multi-user interference and the second term 
the noise variance at the $k^{th}$ GT. Vector $\vb*{w}_{k} \in\mathbb{C}^{M \times 1}$ is the beamforming vector of the $k^{th}$ GT applied at the GBS. The GBS beamforming matrix $\vb*{W} \in \mathbb{C}^{M \times K}$ contains $K$ beamforming vectors
, 
$\vb*{W} =[\vb*{w}_{1}, \cdots, \vb*{w}_{K}]$. 
The allocated transmit power for the $k^{th}$ GT 
is $\parallel\vb*{w}_{k} \parallel^2$.
The achieved sum-rate of the system can then be calculated as 
\begin{equation}
        \rho =   \sum\limits_{k=1}^K log_2 \ (1+ \ \gamma_{k})
    \label{eq:xs}. 
\end{equation}

\subsection{Channel Model}
We consider small-scale Rician fading where a line of sight (LoS) component coexist with non-LoS (NLoS) components for modeling the air-to-ground communications channel between the ARIS and the GBS as well as between the ARIS and the GTs~\cite{RISARIS}. 

Equation 
\begin{equation}
\vb*{G}_{TR}=\frac{\sqrt{\lambda_0}}{D^{\alpha}_{TR}}\bigg( \sqrt{\frac{\beta}{1+\beta}} \ \vb*{G^{LoS}_{TR}} + \sqrt{\frac{1}{\beta+1}} \ \vb*{G^{NLoS}_{TR}} \bigg) 
\end{equation}
models the the MIMO communications channel between the $M$ antennas of the GBS and the $N$ passive reflecting elements of the ARIS, where $\lambda_0$ is the path loss at the reference distance of $1 \ m$, $D_{TR}$ is the 3D distance between the GBS and the ARIS
, $\alpha$ is the path loss exponent, $\beta$ is the Rician factor, and $\vb*{G^{LoS}_{TR}}$ and $\vb*{G^{NLoS}_{TR}}$ are the LoS and NLoS components. 
Without loss of generality, the entries of $\vb*{G^{NLoS}_{TR}}$ 
are independent and identically distributed (i.i.d.) and are modeled as zero mean and unit variance circularly symmetric complex Gaussian (CSCG) variables, 
$\sim \mathcal{CN}(0,1)$. 

The LoS channel component can be expressed as
\begin{equation}
\vb*{G}^{LoS}_{TR}=\vb*{G^{(A)}_{TR}} \ \vb*{G^{(D)}_{TR}}, 
\end{equation}
where $\vb*{G^{(D)}_{TR}}$ and $\vb*{G^{(A)}_{TR}}$ correspond to the channel contributions resulting from the angel of departure (AoD) at the GBS and the angel of arrival (AoA) at 
the ARIS. 

The AoD channel contribution is
\begin{equation}
\vb*{G^{(D)}_{TR}}=\Big[ 1, e^{-j\frac{2\pi}{\lambda}\Upsilon\Gamma^{TR}}, \cdots, e^{-j\frac{2\pi}{\lambda}(M-1)\Upsilon\Gamma^{TR}} \Big], 
\end{equation}
where $\lambda$ is the carrier wavelength, $\Upsilon$ is the antenna separation, and $\Gamma^{TR}$ is the AoD component of the transmitted signal from the GBS to the ARIS. The AoD component can be written as $\Omega^{TR}= sin \ \vartheta \ cos \ \psi$, with $\vartheta$ being the elevation AoD and $\psi$ the azimuth AoD from the ULA of the GBS. 

The 
AoA can be calculated as
\begin{equation}
\vb*{G^{(A)}_{TR}}=\Big[ 1, e^{-j\frac{2\pi}{\lambda}\Upsilon\Lambda^{TR}}, \cdots, e^{-j\frac{2\pi}{\lambda}(N-1)\Upsilon\Lambda^{TR}} \Big], 
\end{equation}
where 
$\Lambda^{TR}$ is the AoA component of the transmitted signal from the GBS to the ARIS. This AoA component is given by $\Lambda^{TR}= cos \ \Theta \ sin \ \varphi$, where $\Theta$ is the azimuth AoA and $\varphi$ is the elevation AoA. 


The channel between the ARIS and the GTs, 
\begin{equation}
\vb*{g}_{RK}=\frac{\sqrt{\lambda_0}}{D^{\alpha}_{RK}}\bigg( \sqrt{\frac{\beta}{1+\beta}} \ \vb*{g^{LoS}_{RK}} + \sqrt{\frac{1}{\beta+1}} \ \vb*{g^{NLoS}_{RK}} \bigg), 
\end{equation}
is a function of the distance based path loss and the LoS and NLoS channel gains. 
The $\vb*{g^{NLoS}_{RK}}$ entries follow the same CSCG distribution that we defined earlier for the $G^{NLoS}_{TR}$. 
Since the ARIS communicates with the single-antenna GTs over a MISO communications link, the $\vb*{g^{LoS}_{RK}}$ can be defined as  
\begin{equation}
\vb*{g^{LoS}_{RK}}=\Big[ 1, e^{-j\frac{2\pi}{\lambda}\Upsilon\chi^{RK}}, \cdots, e^{-j\frac{2\pi}{\lambda}(N-1)\Upsilon\chi^{RK}} \Big], 
\end{equation}
where $\chi^{RK} = cos \ \Phi \ sin \ \Omega$ is the the AoD component of the 
signal departing from the ARIS,  
with $\Phi$ being the azimuth AoD and $\Omega$ the elevation AoD. 

\subsection{Problem formulation}
Our objective is to maximize the achievable downlink sum-rate of the MU-MISO communications system by jointly optimizing the active beamforming at the GBS and the passive beamforming 
at the ARIS. The optimization problem is subject to the GBS power constraint and the ARIS phase shift constraint. The power allocation  for each transmission 
must be less than the 
maximum allowed transmission power 
$P_{max}$. The transmission power constraint can be formulated as 
\begin{equation}
    tr\big( \mathbb{E}\big[\vb*{W} \ (\vb*{W})^H \big]\big) \leq P_{max}.
\end{equation}

{The ARIS phase shifts are constrained by} {
{the unit modulus constraint, $\mid  e^{j\theta_{n}} \mid = 1,$ that applies to all $N$ reflecting elements,} 
where $\theta_{n} \in [0, 2\pi), n \in [1,2,...,N]$.} This establishes the ARIS as a steerable, non-amplifying reflector. 

The optimization problem can then be formulated as
\begin{equation}\label{eq:optmzn_1}
\begin{aligned}
& \underset{\vb*{W}, \vb*{\phi}  }{\text{max}} \ \
\rho \\
& \text{s.t.} \ \  tr\big( \mathbb{E}\big[\vb*{W} \ (\vb*{W})^H \big]\big) \leq P_{max},\\
& \ \ \ \ \ \mid  e^{j\theta_{n}} \mid = 1, \ \forall{n=[1, \cdots, N]}.
\end{aligned}
\end{equation}

{
The objective function of~(\ref{eq:optmzn_1}) is non-concave over $\vb*{W}$ and $\vb*{\phi}$. 
The unit modulus constraint of $\vb*{\phi}$ has been shown to be non-convex~\cite{zhang}. Therefore, the optimization problem is found to be a non-convex and non-trivial optimization problem.} 

\section{Proposed Solution} 
\vspace{-1 mm}
\label{sec:solution}
{Since there is no standard method for solving such a non-convex optimization problem, 
we investigate 
data-driven 
solutions 
instead of applying conventional mathematical optimization tools. Most of the traditional solutions to equivalent multi-parameter optimization problems 
are iterative, 
{alternately} optimize the parameters, and reach suboptimal results~\cite{TVT}.}

We propose jointly determining the GBS beamforming matrix and the ARIS phase shifts 
by applying a transition process based on the current state of the system. Since the next system state is independent of the previous states and actions, the process can be modeled as a Markov decision process (MDP). This facilitates applying a RL algorithm without requiring the knowledge of the system model to find the optimal $\vb*{W}$ and $\vb*{\phi}$.


In what follows, we first describe the deep RL (DRL) model and 
define the states, the actions, and the reward. Then, we introduce 
the proposed DDPG algorithm for the DRL model to maximize the sum-rate of the system. 
The DDPG employs two DNNs---the actor network and the critic network---to avoid intractably high dimensionality for the high state-action space. 

\subsection{Deep Reinforcement Learning Model} 
The MDP for the RL agent is composed of the state space $\mathcal{S}$, the action space $\mathcal{A}$, the reward space $\mathcal{R}$, and the transition probability space $\mathcal{T}$, i.e., $\mathcal{(S, A, R, T)}$. At time slot $t$, the agent observes the state $s_t \in \mathcal{S}$, and based on its policy, takes an action $a_t \in \mathcal{A}$. Depending on the distribution of the transition probability $\mathcal{T}(s_{t+1}|s_t, a_t)$, the agent transfers the system to the new state $s_{t+1}$.  
Since the transition probability is highly dependent on the 
environment and it is difficult to obtain, we employ RL, where the agent reaches an optimal action by observing the instant reward and the state transitions while interacting with the environment through the agent's actions. This means that we do not need to know $\mathcal{T}$, but rather need to carefully define the states, the actions, and the reward. 

\textbf{State:} The set of states contains the different observations representing the environment and can be defined as $\mathcal{S} = \{s_1, s_2, ..., s_t, .., s_T\}$. Each state $s_t$ at time slot $t$ 
captures the transmission power of the GBS, the received power of GTs, the previous 
action, and the GBS-ARIS 
and the ARIS-GTs channel matrices. 

\textbf{Action:} {The choices that 
an agent has to 
transition from the current state to the next state comprise the action space.} We define a set of actions that the agent takes as 
$\mathcal{A} = \{a_1, a_2, ..., a_t, .., a_T\}$. Action $a_t$ at time $t$ encapsulates two parts: the active beamforming matrix $\vb*{W}$ of the GBS and the passive beamforming matrix $\vb*{\phi}$ of the ARIS. 

\textbf{Reward:} After taking an action $a_t$ in state $s_t$ at time $t$, the agent receives a reward $R_t(s_t, a_t)$. 
This reward 
is an evaluation metric 
of the taken action. 
The agent is rewarded positively for actions that may lead to a desired goal such as the increase of system performance, 
whereas 
the agent is penalized if the taken actions are counterproductive. 
For our system, we define the reward function 
as the instantaneous sum-rate of the 
MU-MISO downlink, i.e., ${R_t(s_t, a_t)} = \rho$.

The objective is to find the optimal policy $\mathcal{\Tilde{\pi}}$ that maximizes the cumulative discounted reward, where policy $\mathcal{\pi}$ is a function 
that specifies what action to take in each state. The cumulative discounted reward can be written as $\sum\limits_{t \geq 0} \zeta^t R_t $, where $\zeta^t \in [0, 1]$ is the discount factor in the $t^{th}$ time slot that affects the importance of the future reward. The optimal policy $\mathcal{\Tilde{\pi}}$ can thus be calculated as   
\begin{equation}
\mathcal{\Tilde{\pi}} = \underset{\pi} {arg \ max} \ \mathbb{E} \ \Bigg[\sum\limits_{t \geq 0} \zeta^t R_t \mid \pi \Bigg].
\label{equ:policy}
\end{equation}

Q-learning is a model-free algorithm that is commonly 
applied to find the optimal policy for the state-action relationship. The Q-value function $Q(s,a)$ is used to examine the quality of a state-action pair as a function of the expected cumulative reward under policy $\pi$. 
It allows evaluating the transition between states and the actions taken by the agent. 
It Q values 
are stored in the Q-table consisting of the environmental states 
in the rows and the possible
actions of the agent in the columns. The 
table is 
initially filled with random numbers. The Bellman equation is then used to obtain the optimal state-action pairs 
\cite{Survey17deep}. 
The Q-function that fulfills the Bellman equation can be written as
\begin{flalign}
Q^{*}(s, a) = E_{s^\prime}\bigg[R(s, a) + \zeta \times \max_{a\in \mathcal{A}} \ Q(s^\prime, a^\prime)\bigg],
 \label{eq:Bellman}
 \end{flalign} 
where the $s^\prime$ and $a^\prime$ symbolize the next state and action. The Bellman equation is used through a recursive process to update the Q-values until reaching the optimal Q-value function. The update of the Q-function is formulated as   
\begin{flalign}
Q^{*}(s, a) \gets \ &(1-\wp) \ Q^{*}(s, a) + \wp  \\ &\bigg[R(s, a) + \zeta \times \max_{a\in \mathcal{A}} \ Q(s^\prime, a^\prime)\bigg],
 \label{eq:Qupdate}
 \end{flalign} 
 where $\wp$ is the learning rate that is employed to determine how quickly an agent abandons the previous Q-value for the new Q-value. 

Because of scalability limitations and computational infeasibility of calculating every state-action pair in high dimensional state and action spaces of most real-life problems, Q-learning may not be the optimal choice for converging 
to the optimal $Q^*(s,a)$. Therefore, and alternative to the 
 time-consuming and impractical use of Q-learning, 
 it is possible to estimate $Q(s,a)$ using a function approximator. 
 
 Deep neural networks (DNN) 
 can be used as approximators for determining the state-action pairs and policy. It is still possible to converge to a local optimal solution when applying a DNN because of the non-stationary targets and the correlation between the input samples in the time domain. For tackling this problem and breaking the possible correlations between sequential states of the environment, 
 the experience replay is implemented 
 to introduce randomness. 
 When an action is taken by the agent, the system generates a record of experience. At time step $t$, the experience consists of the current state $s_t$, the action $a_t$, the reward $r_t$, and the next state $s_{t+1}$, forming the tuple $e_t = (s_t, a_t, r_t, s_{t+1})$. Each such experience is buffered in a replay memory with the capacity of $\aleph$, such that $\mathcal{M}=\{e_1, ..., e_t, ..., e_{\aleph} \}$. The update of the DRL can now be performed by feeding the mini-batch samples from the replay memory instead of from the last state.     
 \begin{figure}[t]
  \centering
  \includegraphics[height=5 cm, width= 8.65 cm]{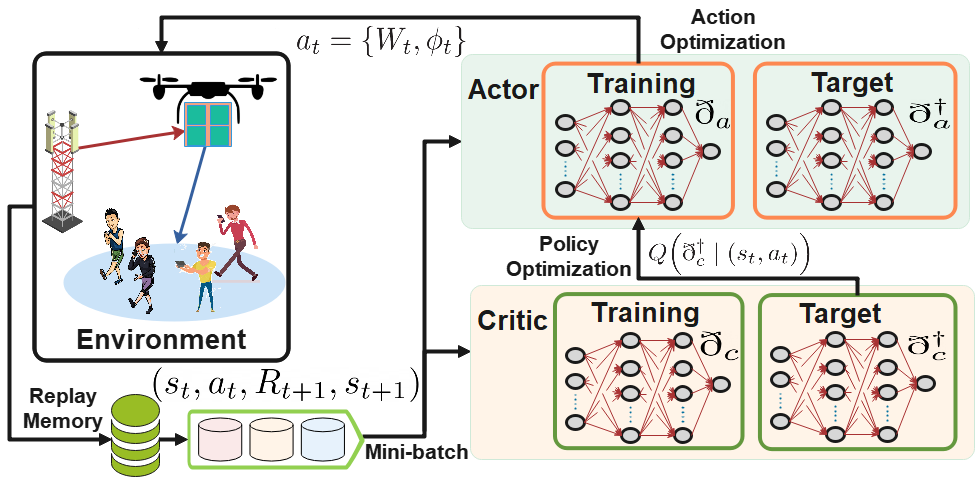}
  \caption{Block diagram of the proposed DDPG architecture.}
  \label{fig:DQL}
\end{figure}

The Q-values of the DRL algorithm are 
$Q(\eth \mid (s,a))$, where $\eth$ corresponds to the weights and the biases 
of the neural networks that are implemented as part of the DRL algorithm. To reach the best estimate for the Q-values, the DRL needs to find the optimal weights and biases 
for minimizing the loss function, which is done using the stochastic gradient descent approach. The update of $\eth$ is thus calculated as      
\begin{flalign}
\eth = \eth - \varrho \ \Delta_{\eth} \ \ell (\eth),
 \label{eq:SGA}
 \end{flalign} 
where $\varrho$ is the learning rate of the stochastic gradient descent updates and $\Delta_{\eth}$ is the gradient of the loss function with respect to $\eth$. 
Parameter $\ell(\eth)$ is the loss function that is calculated as the difference between the estimated output of the neural network and the actual output, that is, 
\begin{flalign}
\notag \ell(\eth) =  \mathbb{E} \Bigg[ \bigg( \Big[ R_{t} + \zeta \times \max_{a\in \mathcal{A}} \ Q \Big( \eth^{\dagger} \mid (s_{t+1}, a_{t+1}) \Big)\Big] - \\ 
\Big[Q \Big( \eth \mid (s_{t}, a_{t}) \Big) \Big] \bigg)^2 \Bigg],
\label{eq:loss}
\end{flalign}
where $ Q \Big( \eth^{\dagger} \mid (s_{t+1}, a_{t+1}) \Big)$ is the target Q-value with $\eth^{\dagger}$ target 
coefficients and $Q \Big( \eth \mid (s_{t}, a_{t}) \Big)$ is the actual Q-value. To produce both actual and target Q-values, the DRL implements two identical neural networks known as the training neural network and the target neural network. The training network outputs the Q-values associated with the actions of the agent in each state. The target network supervises the training network by providing the target Q-values obtained from the Bellman equation.   

\subsection{Deep Deterministic Policy Gradient}
DRL can 
be effective when applied to discrete action outputs, but not for 
continuous action spaces. 
We therefore propose to employ the DDPG~\cite{DDPG} for continuous decision‐making by efficiently learning and acting in the continuous action spaces of the active and passive beamforming systems of 
Fig. 1. 

The DDPG relies on two types of DNNs---the actor and the critic networks, each with its training and target network (Fig.~\ref{fig:DQL})---to avoid the intractably high dimensionality of the state-action space. 
It develops the mapping function between the input environment states and the agent actions to establish a deterministic policy that maximizes the long-term reward. 
The actor network is adopted to generate the 
deterministic policy that maximizes the output of the critic network; that is, it takes the states as the input and outputs the deterministic actions to the agent. 
The critic network is designed to simulate the Q-value function and evaluate the deterministic policy generated by the actor network; that is, it takes the deterministic policy of the actor network as its input and outputs the Q-value for these actions. 

The updates of the training critic network are obtained as
\begin{flalign}
\eth_{c} = \eth_{c} - \varrho_{c} \ \Delta_{\eth_{c}} \ell (\eth_{c}),
 \label{eq:CriticSGA}
 \end{flalign}
 where $\eth_{c}$ 
 captures the weights and the bias of the training critic network, $\varrho_{c}$ is the learning rate, 
 and $\Delta_{\eth^{t}_{c}}$ is the gradient. 
 Parameter $ \ell (\eth_{c})$ is the loss function of the training critic network and can be calculated as 
 \begin{flalign}
\notag \ell (\eth_{c}) =  \mathbb{E} \Bigg[ \bigg( \Big[ R_{t} + \zeta \times \ Q\Big( \eth_{c}^{\dagger} \mid (s_{t+1}, \Tilde{a}) \Big)\Big] - \\ 
\Big[Q \Big( \eth_{c} \mid (s_{t}, a_{t}) \Big) \Big] \bigg)^2 \Bigg],
\label{eq:lossCritic}
\end{flalign}
where $\Tilde{a}$ is the action of the agent that follows the deterministic policy drafted by the target actor network and $\eth_{c}^{\dagger}$ captures the network's weights and bias. 

It is 
worth mentioning that the update of the training network occurs more frequently than the update of the target network. 
The training actor network update is defined as 
\begin{flalign}
\eth_{a} = \eth_{a} - \varrho_{a} \ \Delta_{a}Q \Big( \eth_{c}^{\dagger} \mid (s_{t}, a_{t}) \Big) \ \Delta_{\eth_{a}}\mho(\eth_{a} \mid s_{t}),
 \label{eq:ActorSGA}
 \end{flalign}
 where $\eth_{a}$ 
 corresponds to the weights and bias of the training actor network $\mho(\eth_{a} \mid s_{t})$, $\varrho_{a}$ is the learning rate, 
 $\Delta_{a}Q \Big( \eth_{c}^{\dagger} \mid (s_{t}, a_{t}) \Big)$ is the gradient of the target critic network output with reference to the taken action, and $\Delta_{\eth_{a}} \ \mho(\eth_{a} \mid s_{t})$ is the gradient of the training actor network with respect to $\eth_{a}$. The gradient $\Delta_{a}Q \Big( \eth_{c}^{\dagger} \mid (s_{t}, a_{t}) \Big)$ is introduced in the update of the training actor network to guarantee that the upcoming selection of the action by the critic network is the preferred one for maximizing the Q-value function.  
 
 \begin{minipage}{0.95\linewidth}
\begin{algorithm}[H]
\footnotesize
\SetAlgoLined
\nl \text{\bf{Input}} $\vb*{G}_{TR}$, $\vb*{g}_{RK}$\\
\nl \text{\bf{Initialize}} \ $\vb*{W}$ and $\vb*{\phi}$\\
\nl \text{\bf{Initialize}} \ $\mathcal{M}$ with capacity $\aleph$,\ $\zeta$, $\eth_{c}$, $\eth_{c}^{\dagger}=\eth_{c}$, $\varrho_{c}$, $\eth_{a}$, $\eth_{a}^{\dagger}=\eth_{a}$,\ $\varrho_{a}$

\nl \For{episode = 1, 2, ..., E}
{    
\nl    \it{Obtain state $s_1$} \\
\nl     \For{t = 1, 2, ..., T}
    {
\nl        From the training actor network, acquire $ a_t = \{\vb*{W_t}, \vb*{\phi_t} \}$\\
\nl         Observe next state $s_{t+1}$ and instant reward $R_{t+1}$ given $a_t$ \\ 
\nl         Store experience $e_t = (s_t, a_t, R_{t}, s_{t+1})$ in $\mathcal{M}$\\
\nl         Obtain $Q \Big( \eth_{c} \mid (s_{t}, a_{t}) \Big)$ from the training critic network\\
\nl         Calculate $\ell (\eth_{c}) $ via eq.~(\ref{eq:lossCritic})\\
\nl         Calculate $\Delta_{\eth_{c}} \ell (\eth_{c})$, $\Delta_{a}Q \Big( \eth_{c}^{\dagger} \mid (s_{t}, a_{t})\Big)$, $\Delta_{\eth_{a}}\mho(\eth_{a} \mid s_{t})$ \\
\nl         Update critic and actor training networks $\eth_{c}$ and $\eth_{a}$\\
\nl         Update critic and actor target networks $\eth_{c}^{\dagger}$ and $\eth_{a}^{\dagger}$ after $\varepsilon$ steps\\
\nl        Train the DNN network with $s_{t+1}$ as input \\
    }
 }
\KwResult{Optimal $\vb*{W}$ and $\vb*{\phi}$}
\caption{\small DDPG-based Joint Active and Passive Beamforming Design.}
\label{Algorithm1}
\end{algorithm}
\end{minipage}
\\
\\

 Coupled with the updates of the training critic and training actor networks, the target critic and target actor network updates are defined as
 \begin{flalign}
&\eth_{c}^{\dagger} \gets \eta_{c} \ \eth_{c} + (1 - \eta_{c}) \ \eth_{c},\\
&\eth_{a}^{\dagger} \gets \eta_{a} \ \eth_{a} + (1 - \eta_{a}) \ \eth_{a},
 \label{eq:TargetCriticSGA}
 \end{flalign}
where $\eta_c$ and $\eta_a$ are the learning rates for updating 
the critic and actor networks, respectively. The details of the proposed DDPG solution are presented in Algorithm~\ref{Algorithm1}. \textcolor{black}{The DDPG 
can be integrated into radio access network intelligent controllers (RICs) where the training can be done at the non-real time-RIC and the inference at the near-real time RIC of future O-RAN systems~\cite{abdalla2021toward}.}

\section{Numerical Analysis and Discussion}
\vspace{-1 mm}
\label{sec:results}
  \begin{figure}[t]
  \vspace{+0.3 mm}
     \centering
     \includegraphics[width=0.5\textwidth]{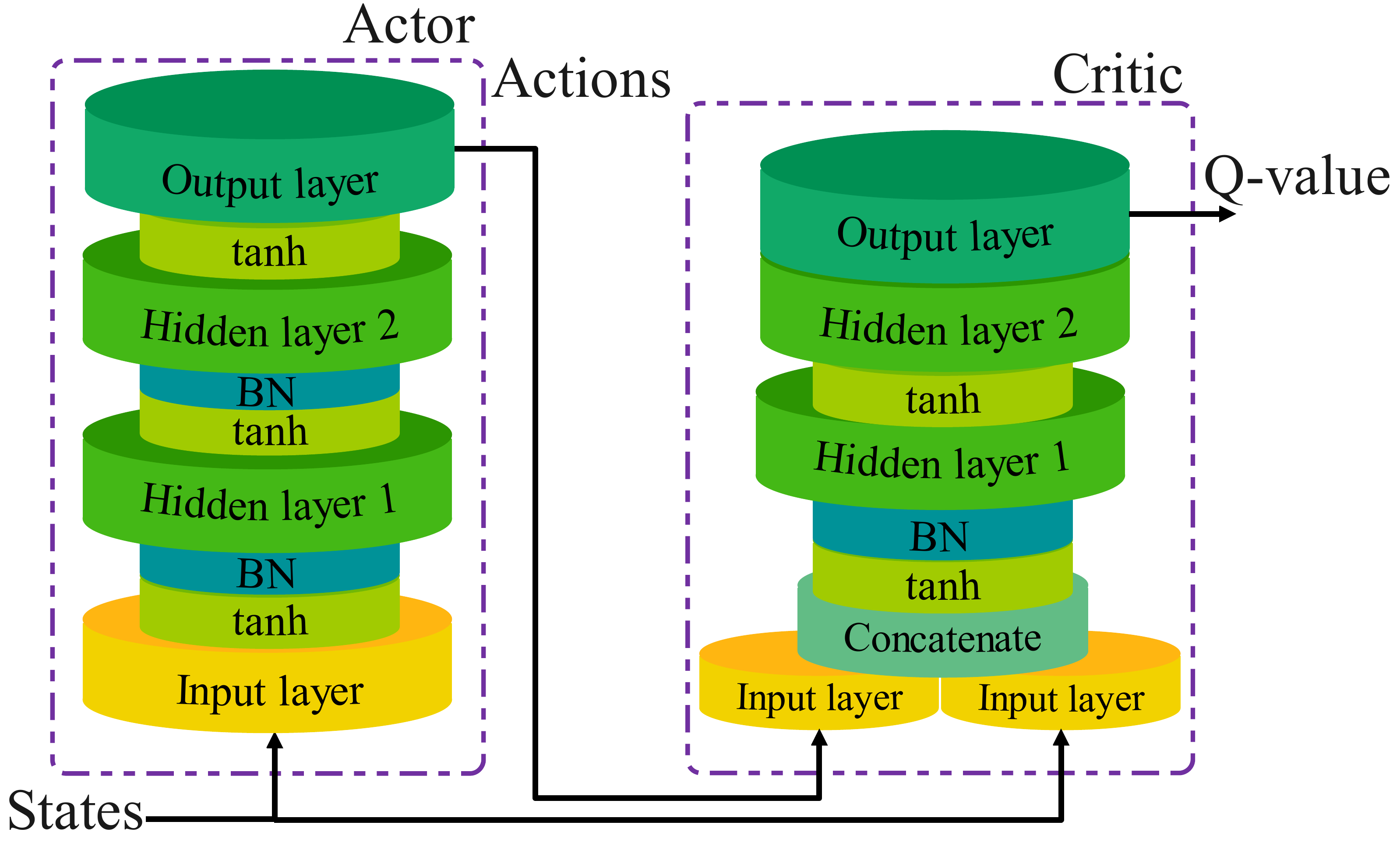}
     \caption{The DNN design for the actor and critic networks used in the proposed DDPG algorithm (BN--batch normalization).}
     \vspace{-5 mm}
     \label{fig:DNNsArch}
 \end{figure}
 \begin{figure}[b]
 \vspace{-6 mm}
     \centering
     \includegraphics[width=0.48\textwidth]{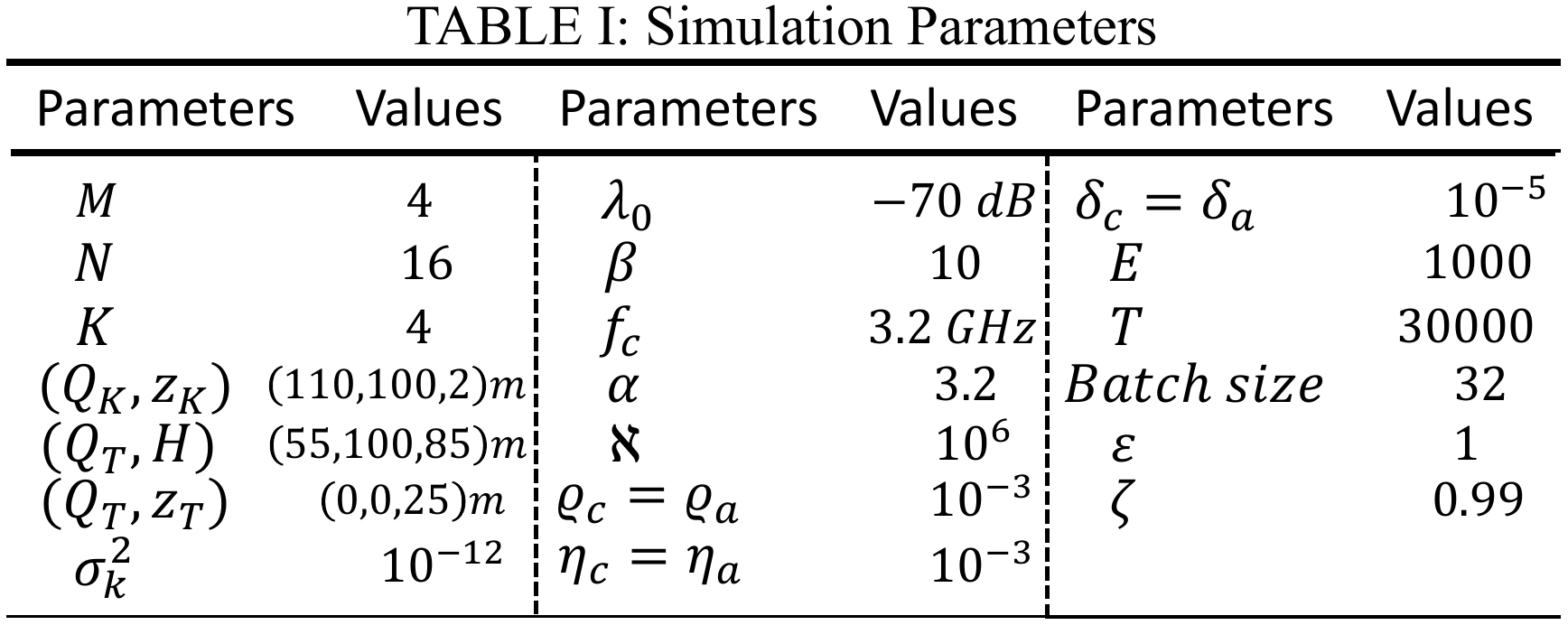}
     \label{fig:Table}
 \end{figure}
 
We numerically analyze the performance of the proposed 
scheme for optimizing the downlink sum-rate of the ARIS assisted MU-MISO system. 
The GTs are randomly distributed in a 2D circular area that is centered at $Q_K$. 
Table I provides the parameters for the simulated scenarios and the hyperparameters for the proposed DDPG algorithm. 
The critic and actor DNNs employ the same structure and consist of $4$ 
layers: the input layer with $N_i^{\prime}$ neurons, two fully connected {hidden} layers with $532$ neurons each
, and the output layer with $N_o^{\prime}$ neurons. Parameter $N_i^{\prime}=2K+2KN+2K^2+2N+2NM+2MK$ corresponds to the size of the state space and $N_o^{\prime}=2N+2MK$ corresponds to the size of the action space. All DNNs employ $tanh$ as the activation function in all layers and $Adam$ with adaptive learning as the optimizer, where $\varrho_c^{t+1}=\delta_c\varrho_c^{t}$ and $\varrho_a^{t+1}=\delta_a\varrho_a^{t}$ with the decaying rates $\delta_c$ an $\delta_a$ for the critic and actor networks, respectively. {Fig.~\ref{fig:DNNsArch} illustrates the structure of the critic and actor DNN designs for the proposed DDPG.}  

    \begin{figure*}
  \centering
  \begin{subfigure}[b]{0.32\textwidth}
    \centering
    \includegraphics[width=\textwidth]{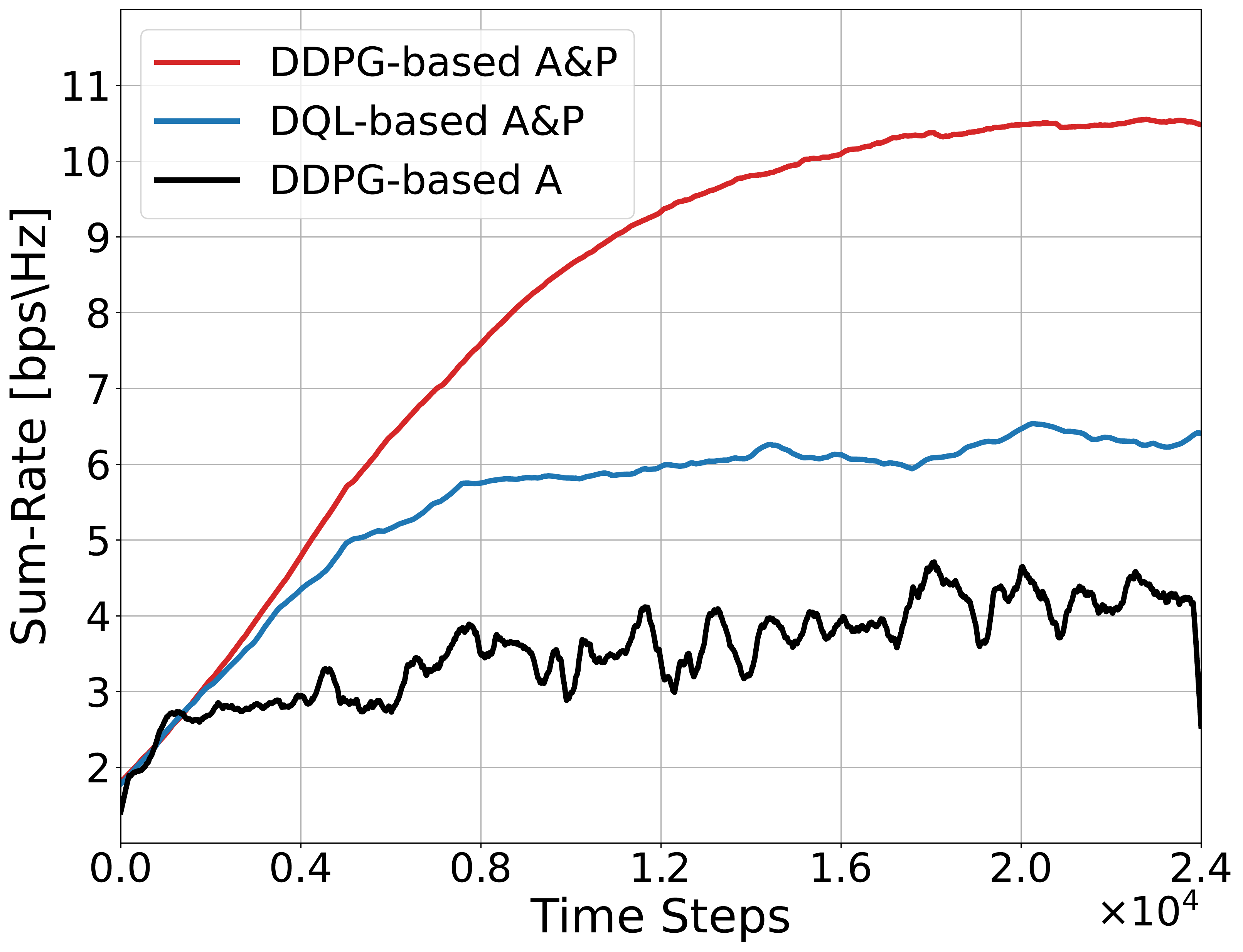}
    \caption{}
    \label{fig:Vs}
  \end{subfigure}
  \begin{subfigure}[b]{0.32\textwidth}
    \centering
    \includegraphics[width=\textwidth]{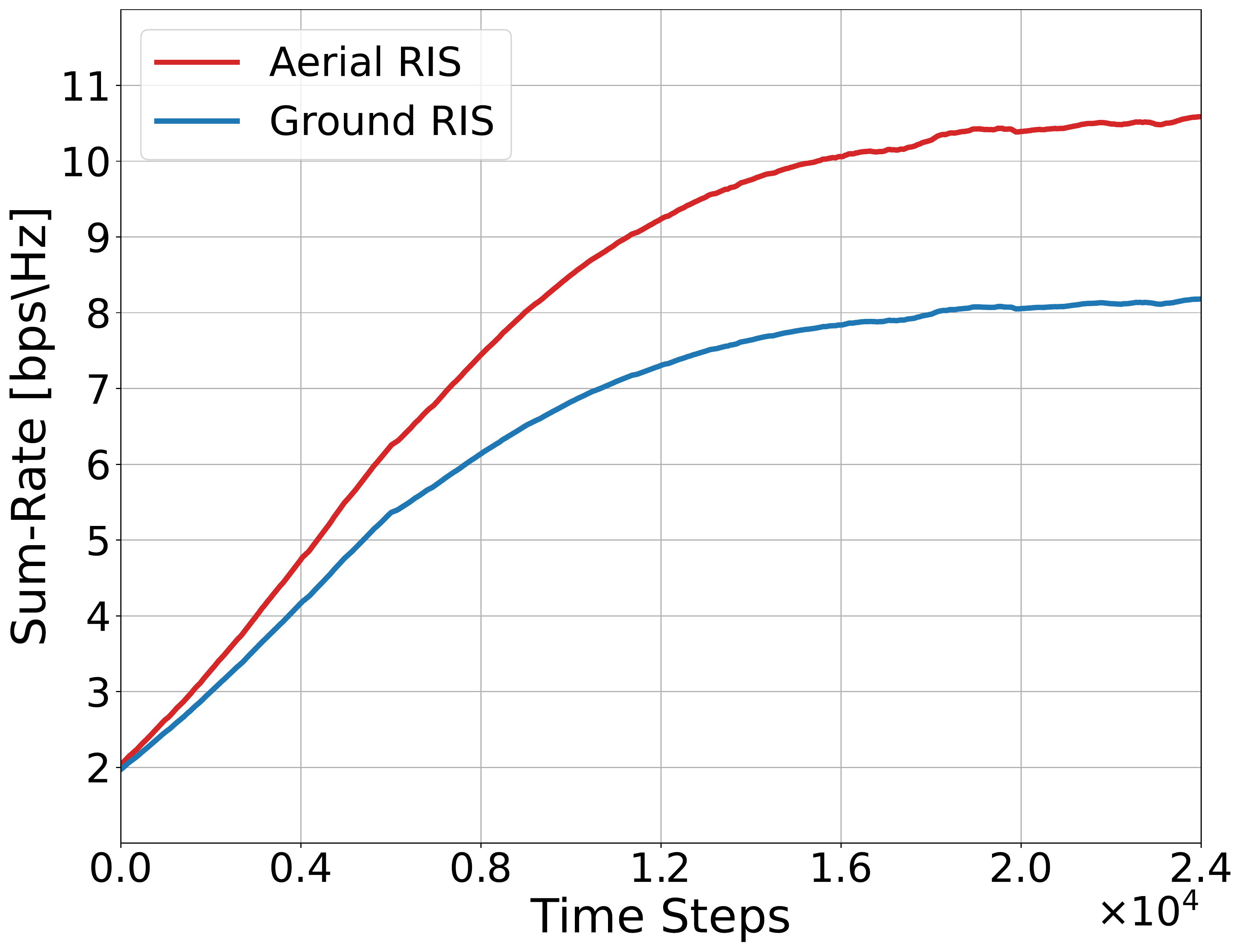}
    \caption{}
    \label{fig:Distance}
  \end{subfigure}
  \begin{subfigure}[b]{0.32\textwidth}
    \centering
    \includegraphics[width=\textwidth]{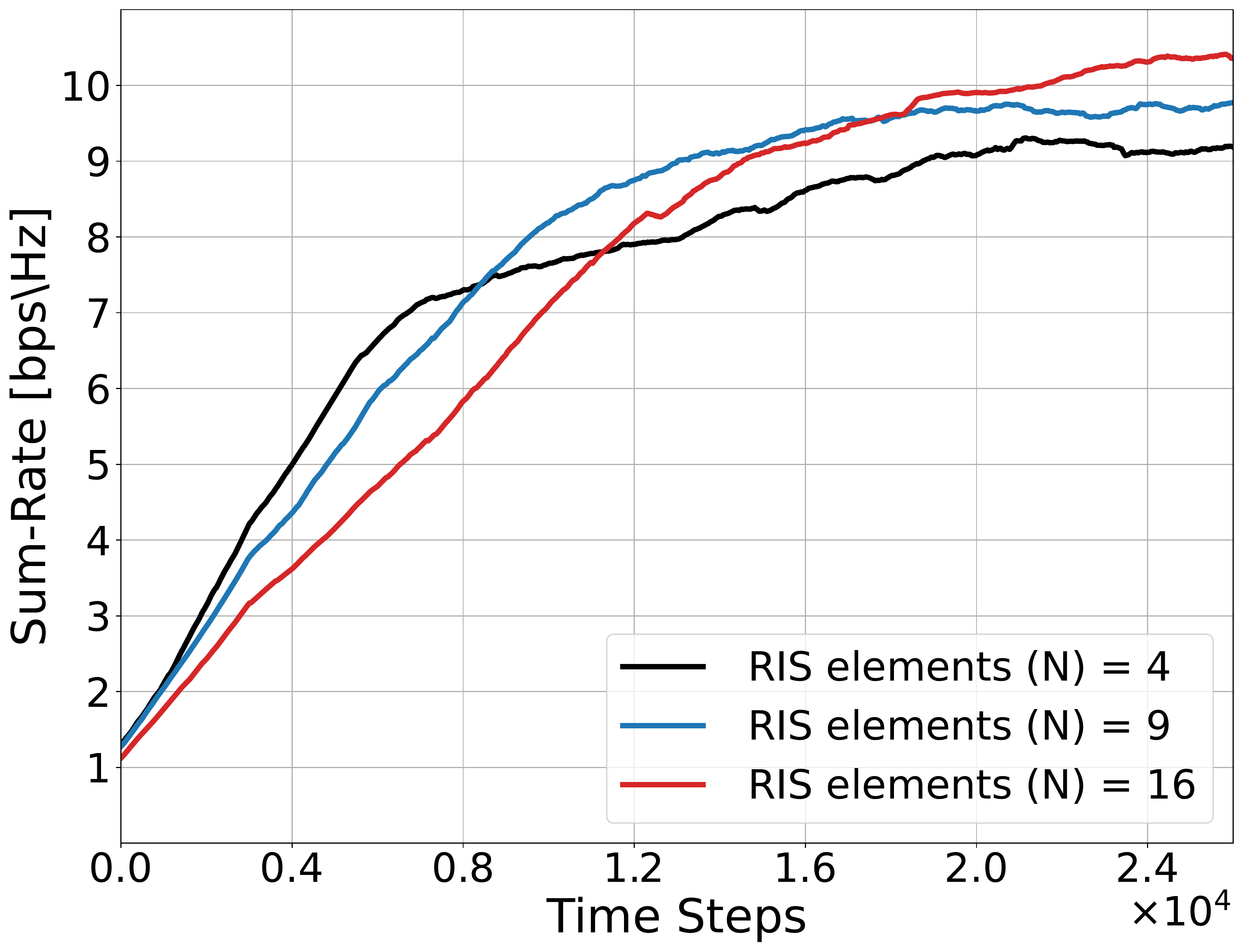}
    \caption{}
    \label{fig:RISElements}
  \end{subfigure}
  \vspace{-2mm}
  \caption{Achieved sum-rate comparison for the DDPG, DQL, and random phases (a), for the aerial RIS and ground RIS (b), and for different ARIS elements $N$ (c). }
  \label{fig:Results}
  \vspace{-5mm}
\end{figure*}

 
For the performance evaluation, we compare the resulting reward of the DDPG-based algorithm for optimizing the joint transmit beamforming and phase shifts of ARIS (\textit{DDPG-based A\&P} for active and passive beamforming optimization) against a deep Q-learning (DQL)-based approach (\textit{DQL-based A\&P}) which uses the same objective function and the same reward function as the proposed DDPG algorithm. We also compare our solution to a baseline technique (\textit{DDPG-based A}) that only optimizes the GBS beamforming matrix and chooses the ARIS phase shifts randomly.
The DQL-based solution uses the same DNN structure used for the critic and the actor networks. 

Figure~\ref{fig:Results}.a illustrates the achieved sum-rates of the MU-MISO downlinks over time in discrete time slots for the three optimization methods. 
The sum-rate corresponds to the reward function.  
Overall we notice a significant increase in the sum-rate of the proposed DDPG-based algorithm over the DQL-based algorithm and the baseline. 
The superiority of the DDPG over the baseline applying random phases 
can be attributed to the RIS being able to influence the propagation environment in such a way to suppress the multi-user interference. 
The 
superiority of the DDPG over the DQL solution comes from the {actor-critic relationship}---{described in the previous section}---that is employed for optimizing the policy that reflects on the agent's action choices. 
The DQL's sum-rate fluctuates steadily around $6 \ bps / Hz$ for most of the training time. The achieved sum-rate of the DDPG converges to roughly 1.75x the 
sum-rate of the DQL. This indicates 
the quality of the actions generated by the DDPG, whereas the DQL converges to a local minimum. 

\textcolor{black}{To provide more insights into the performance of the proposed ARIS-assisted MU-MISO downlink system, Fig~\ref{fig:Results}.b presents a comparison of the achieved sum-rate of utilizing the ARIS versus a ground RIS, where the RIS is deployed on a ground building. The ARIS scheme outperforms the ground RIS, achieving a 30\% higher 
sum-rate. 
This can be attributed to the stronger LoS links that are achievable by the ARIS and that empower the A2G communication channels over the equivalent ground channels. 
}    

Figure~\ref{fig:Results}.c shows the sum-rate of the DDPG algorithm as a function of the number of reflecting elements. 
As expected, 
an ARIS with more elements converges to a higher sum-rate, but it takes longer to converge because of the nature of the problem that has a larger number of states and actions. The 
sum-rate improves by 
$15.4\%$ when employing an ARIS with 16 over 4 reflecting elements, by $8.5\%$ for 9 over 4 elements, and by $6.8\%$ for 16 over 9 elements. 

\section{Conclusions}
\vspace{-1 mm}
\label{sec:conclusions}

This paper has presented a driven-based solution to jointly optimize the active beamforming at the GBS and the passive beamforming at the ARIS for increasing the downlink sum-rate of a MU-MISO system. We have provided the detailed design options and parameters of the proposed DDPG algorithm. We have numerically analyzed the performance of the proposed solution and compared it with the DQL and a baseline technique. The results have shown that the DDPG outperforms other techniques and effectively increases the sum-rate. 
Motivated by these results, we propose 
studying multiple DDPG agents and analyzing the performance of the ARIS-enhanced communications system by optimizing both the beamforming and the location/trajectory of the ARIS.


\section*{\textcolor{black}{Acknowledgment}}
\noindent
This work 
was supported in part by NSF 
award 
CNS-2120442. 

\balance

\bibliographystyle{IEEEtran}
\bibliography{Refs}

\vspace{0.2cm}
\noindent

\vspace{0.2cm}
\noindent

\vspace{0.2cm}
\noindent

\end{document}